\documentclass{ws-p8-50x6-00}

\renewcommand{\be}{\begin{equation}}
\renewcommand{\ee}{\end{equation}}

\newcommand{\ba}{\begin{eqnarray*}}
\newcommand{\ea}{\end{eqnarray*}}
\newcommand{\bi}{\begin{itemize}}
\newcommand{\ei}{\end{itemize}}

\newcommand{\la}[1]{\label{#1}}
\newcommand{\rmi}[1]{{\mbox{\scriptsize #1}}}
\newcommand{\nr}[1]{(\ref{#1})}
\newcommand{\fr}[2]{{\frac{#1}{#2}}}

\newcommand{\eq}{Eq.~}

\newcommand{\fig}{Fig.~}

\def\RR{{\rm I\kern -.2em  R}}

\def\lsi{\raise0.3ex\hbox{$<$\kern-0.75em\raise-1.1ex\hbox{$\sim$}}}
\def\gsi{\raise0.3ex\hbox{$>$\kern-0.75em\raise-1.1ex\hbox{$\sim$}}}
\newcommand{\lsim}{\mathop{\lsi}}
\newcommand{\gsim}{\mathop{\gsi}}

\begin{document}

%\vspace*{-2cm}
%\begin{flushright} hep-ph/0001292
%\end{flushright}
%\vspace*{1cm}

\title{Vortex Phases in \\ Condensed Matter and Cosmology}

\author{M. Laine}

\address{Theory Division, CERN, CH-1211 Geneva 23, Switzerland} % \\
%Dept.\ of Physics, P.O.Box 9, FIN-00014 Univ.\ of Helsinki, Finland} 

\maketitle

\abstracts{Placing a high-$T_c$ superconductor in an increasing 
external magnetic field, the flux first penetrates the sample 
through an Abrikosov vortex lattice, and then a first order 
transition is observed by which the system goes to the normal 
phase. We discuss the cosmological motivation for considering 
the electroweak phase transition in the presence of an external 
magnetic field, the analogies this system might have with the 
superconductor behaviour described above, and in particular 
whether at large physical Higgs masses, corresponding to the 
high-$T_c$ regime, an analogue of the vortex phase and an associated 
first order phase transition could be generated.}

\vspace*{-8cm}
\begin{flushright} CERN-TH/2000-031\\ hep-ph/0001292
\end{flushright}
\vspace*{6.5cm}

\section{Introduction}

One possible explanation for the baryon asymmetry of the Universe
is that it was generated in a 1st order electroweak phase 
transition\cite{krs}. This possibility cannot be realized in 
the Standard Model (SM), though, since there is no 1st order transition
for $m_H\gsim 72$ GeV~\cite{endpoint}. In the MSSM there is still 
room for a strong 1st order transition, if the right-handed
stop is lighter than the top\cite{stop}: 
a consequence of such a circumstance would be a Higgs lighter 
than $\sim 110$ GeV, not yet experimentally excluded for the MSSM.

In this paper, we will consider
another possibility for increasing the strength of the electroweak phase 
transition. Indeed, remaining in the 
SM but imposing an external magnetic field $H_\rmi{ext}$ 
on the system, has a strengthening effect\cite{gs}. It turns out that
for baryon number non-conservation there is an opposing 
effect due to the sphaleron dipole moment\cite{cgpr}, but
we nevertheless consider it interesting 
to map out the phase diagram  in some detail
for $H_\rmi{ext}\neq 0$, as an analogy with superconductors (Sec.~3)
suggests that the system might have quite unexpected properties. The case
of small Higgs masses in $H_\rmi{ext}\neq 0$, 
as well as the first results on large Higgs masses, 
were already reviewed in\cite{sewm}, and we 
concentrate here on the physical case $m_H\gsim m_Z$.

\section{Cosmological motivation for $H_\rmi{ext}\neq 0$}

The physical relevance of considering $H_\rmi{ext}\neq 0$
comes from the observation that the  
existence of galactic magnetic fields today may well imply the 
existence of primordial seed fields in the Early Universe. 
In order to get large enough length scales, it seems conceivable that 
even in the most favourable case of strongly ``helical'' fields,
the seed fields should have a correlation length at least of the 
order of the horizon radius at the electroweak (EW) epoch\cite{son}. 
Such large length scales could possibly be produced 
during the inflationary period of Universe expansion
(see, e.g., ref.\cite{inflation} and references therein).

If a primordial spectrum of magnetic fields is generated
during inflation, it is on the other hand also true that 
after a while the fields are essentially homogeneous
at small length scales. Indeed, magnetohydrodynamics, 
\be
\frac{\partial \vec{H}_Y}{\partial t} = 
\frac{1}{\sigma} \nabla^2 \vec{H}_Y + 
\nabla\times(\vec{v}\times\vec{H}_Y), 
\ee
tells that magnetic fields diffuse away at scales 
$ l \lsim ({t}/{\sigma})^{1/2} \sim 
({M_\rmi{Pl}}/{T})^{1/2} T^{-1}$, 
where $\sigma$ is conductivity.
At the EW epoch $T\sim 100$ GeV, 
this gives $l_\rmi{EW}\sim 10^7/T$, 
a scale much larger than the typical 
correlation lengths $\sim$ a few $\times T^{-1}$. 

A further question is the magnitude of magnetic fields. 
An equipartition argument would say that only a small 
fraction of the total (free) energy density can be in magnetic 
fields. This leads to $H_Y/T^2 \lsim 2$.  
In conclusion, there could well be essentially homogeneous and 
macroscopic (hypercharge) 
magnetic fields around at $T\sim 100$ GeV, with 
a magnitude $H_Y/T^2 \sim 1$.  

\section{Superconductors in $H_\rmi{ext}\neq 0$}

As a further motivation for studying in detail the electroweak case, 
let us recall the very rich structure found in quite an analogous system, 
superconductors
under an external magnetic field. Denoting the inverses of the 
spatial scalar and vector correlation lengths by $m_H,m_W$ (and 
$x\equiv m_H^2/(2 m_W^2)$), the usual starting point for superconductor
studies, the 3d continuum scalar electrodynamics (or the 
Ginzburg-Landau, GL, theory), 
%\be
%{\cal L}_{\rm GL} = 
%{1\over4} F_{ij}F_{ij}+
%(D_i\phi)^* D_i\phi+ y \phi^*\phi+
%x (\phi^*\phi)^2, \la{gl}
%\ee
predicts at the tree-level
two qualitatively different responses of the system 
to an external magnetic field: 

In the type I case, $m_H< m_W$, a flux cannot penetrate 
the superconducting phase. However, superconductivity is destroyed 
by $H_\rmi{ext}$. The way in which this transition has to take
place is that the superconducting and the normal phases 
coexist at $H_\rmi{ext}^c$. This implies a {\em 1st order} transition. 

In the type II case, $m_H>m_W$, on the other hand, the flux can 
penetrate the system via an Abrikosov vortex lattice. At a large
enough $H_\rmi{ext}$ the system then {\em continuously} changes to the 
normal phase.

%%%%%%%%%%%%%%%%%%%%%%%%%%%%%%%%%%% FIGURE %%%%%%%%%%%%%%%%%%%%%%%%
\begin{figure}[t]

\vspace*{-0.5cm}

\centerline{
\epsfxsize=6cm\epsfbox{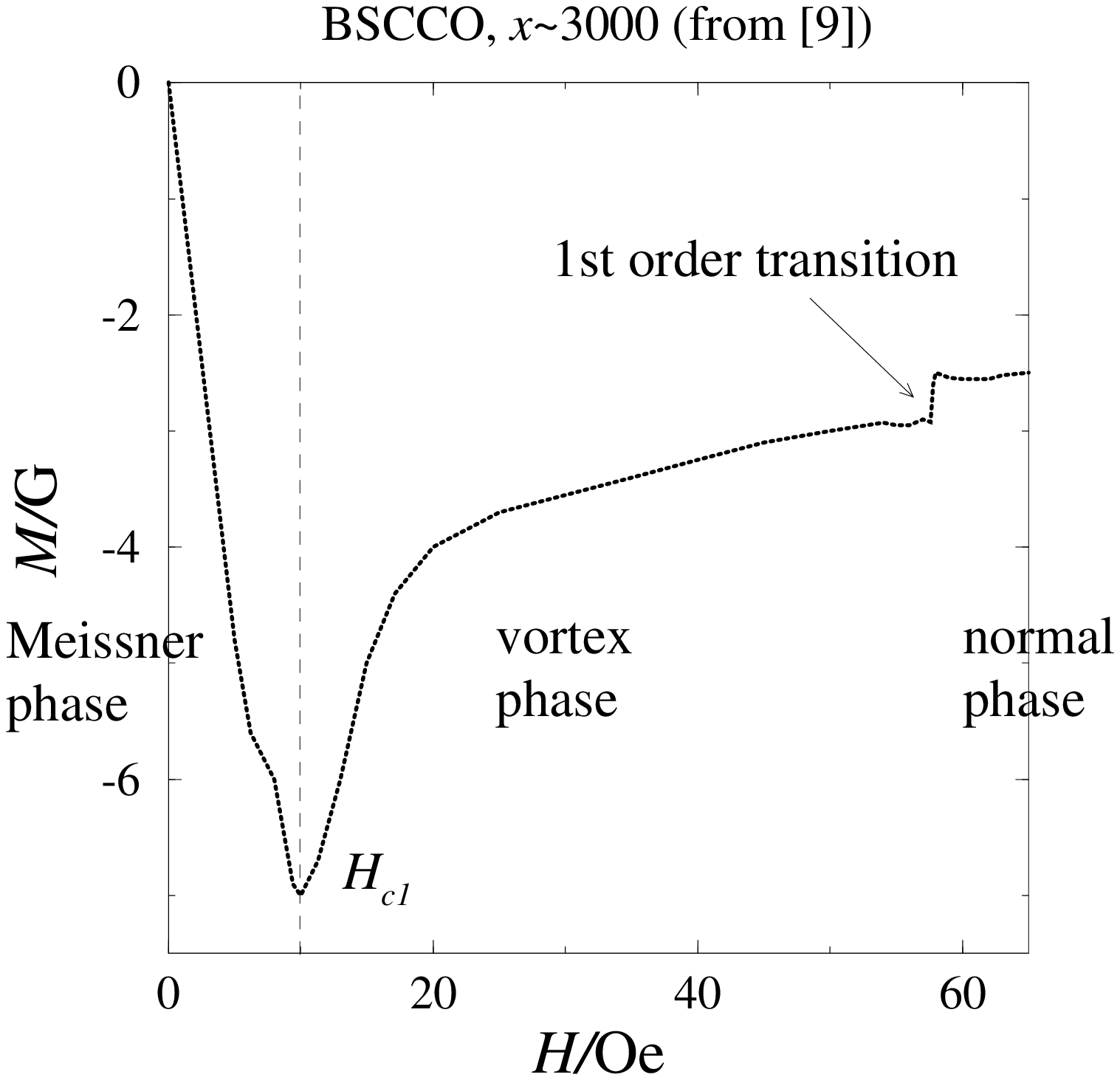}% 
\epsfxsize=6cm\epsfbox{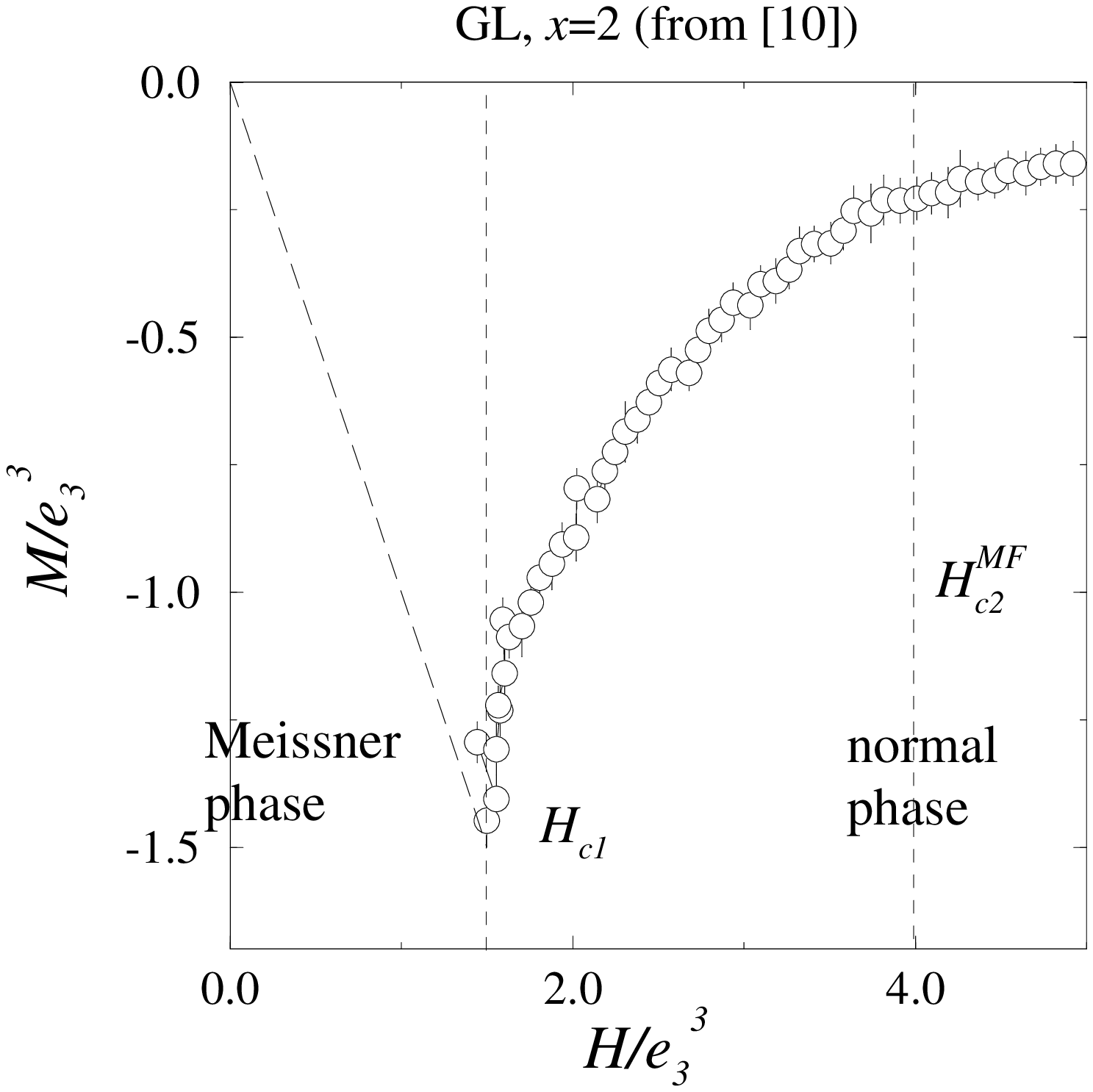}% 
}

\vspace*{-2.5cm}

\caption[]{Left: magnetization $M$ as a function of the field 
strength $H$ as observed for the high-$T_c$ material BSCCO ($x\sim 3000$);
a qualitative reproduction of~Fig.~2 in\cite{nat1}. The tree-level 
(``mean field'')
upper critical field $H_{c2}^\rmi{MF}$ is huge due to the large value of $x$, 
$H_{c2}^\rmi{MF} \sim 10^3 H_{c1}$, and the transition observed takes place
much below $H_{c2}^\rmi{MF}$. 
%($H_{c2}/H_{c1} = 2x/{\cal E}(\sqrt{2x})$, 
%where ${\cal E}(z)$ is a slowly 
%increasing function with ${\cal E}(1)=1$\cite{jr}). 
The system is macroscopic in the sense
that there are several hundred flux quanta at the point of the 
1st order transition. Right: magnetization $M$ as a function of 
$H$ in the GL model at $x=2$; data from\cite{manyvortex}. 
Measurements were only carried out for $H\ge H_{c1}$. 
At $H_{c2}^\rmi{MF}$, 
there are $\sim 40$ flux quanta in the system. 
No 1st order transition is observed within this resolution.
\la{fig:sc}}
\end{figure}
%%%%%%%%%%%%%%%%%%%%%%%%%%%%%%%%%%%%%%%%%%%%%%%%%%%%%%%%%%%%%%%%%%%

It is now a very interesting observation that fluctuations 
change the nature of the tree-level type II transition
described above in an essential way. 
%As was theoretically expected\cite{ffh}, 
Indeed, much of the vortex lattice phase is 
observed to be removed, but it is also found in high-$T_c$ 
superconductors (which 
are strongly of type II) that the continuous transition changes to 
a 1st order one: for a particularly clear signal, 
see Fig.~2 in\cite{nat1}, reproduced in \fig\ref{fig:sc}(left).

At the same time, high-$T_c$ superconductors are of course 
a complicated layered and highly anisotropic material, so it is
not immediately clear whether the 1st order transition observed
is also a property of, say, the simple continuum GL theory.
Let us list arguments in favour of and against this possibility:
\bi
\item
There is an analytic prediction of a 1st order transition\cite{bnt},
starting
just from the GL theory. However, it is based on an $\epsilon$-expansion 
around $d=6$, and relies on $\epsilon=3$ being small. 
Other analytic arguments (see, e.g., \cite{zt})
also lack a small expansion parameter. 
A set of lattice simulations have been carried out which favour the 
possibility of a phase transition directly in the GL theory\cite{ns}.
However, the theory actually simulated is not GL but some approximation
thereof, and moreover, the effects of discretization artifacts in the
simulations have not been systematically investigated. 
\item 
There are, on the other hand, 
other simulation results which argue that a layered
structure {\em is} essential for the 1st order transition\cite{km}. 
However, these simulations use again an approximate form of the
theory, whose 
validity for the full GL model is not clear. 
Finally, direct lattice simulations in the 
full GL model\cite{manyvortex}
have so far failed to see a transition, see \fig\ref{fig:sc}(right). 
However, one can argue that due to the high computational cost, 
these simulations do not necessarily yet represent
the thermodynamical limit with respect to the number of vortices. 
\ei
To summarize, we consider it at the moment an open problem 
what is the ``minimal'' continuum model which may display 
a 1st order transition between the vortex phase and the normal phase.  
Understanding this issue would be very important for, e.g., 
the considerations to which we now turn.

\section{The Electroweak Theory in $H_\rmi{ext}\neq 0$ at Tree-Level}

To analyse the behaviour of the electroweak theory in an 
external magnetic field, we can directly consider the 
dimensionally reduced 3d action\cite{bext} 
\be
{\cal L}_{\rm 3d} = 
{1\over4} G_{ij}^aG_{ij}^a+{1\over4} F_{ij}F_{ij}+
(D_i\phi)^\dagger D_i\phi+y \phi^\dagger\phi+
x (\phi^\dagger\phi)^2. \la{ewaction}
\ee
Here $G^a_{ij},F_{ij}$ are the SU(2) and U$_Y$(1) field strengths, 
and $\phi$ is the Higgs doublet. In terms of the 
physical 4d parameters, $x$ and $y$ are expressed as
\be
x \sim 0.12 \frac{m_H^2}{m_W^2}, 
\quad 
y  \sim 4.5 \frac{T-T_0}{T_0}, 
\ee
where $T_0$ equals the critical temperature up to radiative corrections.
By a magnetic field we now mean, in the symmetric phase of 
the theory, an Abelian U$_Y$(1) magnetic field $H_Y$.
In the broken phase, this goes dynamically to 
the electromagnetic field $H_\rmi{EM}$. 

%In the simulations, what one does is to impose a 
%hyperflux through the system with boundary conditions, 
%$\int dx_1dx_2 F_{12} = \oint d s_i A_i = \Phi_{H_Y}$. 
%The results obtained in an ensemble with a fixed flux can then 
%be converted to an ensemble with a fixed field strength~$H_Y$.

%%%%%%%%%%%%%%%%%%%%%%%%%%%%%%%%% FIGURE
\begin{figure}

\vspace*{-0.7cm}
 
\centerline{
\epsfxsize=6cm\epsfbox{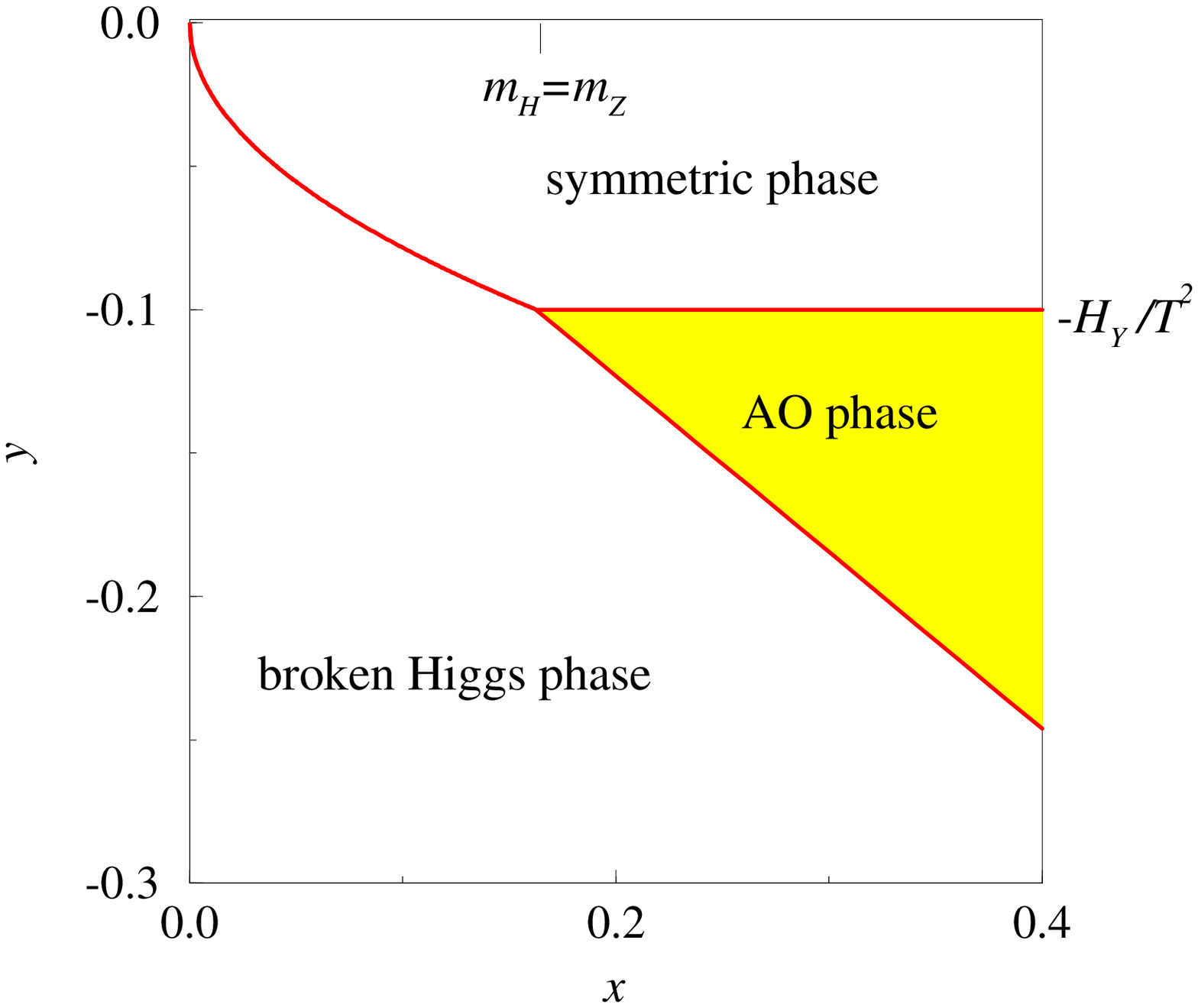}%
\hspace*{5.5cm}
}

\vspace*{-6.4cm}

\centerline{
\hspace*{4.8cm}%
\epsfxsize=4cm\epsfbox{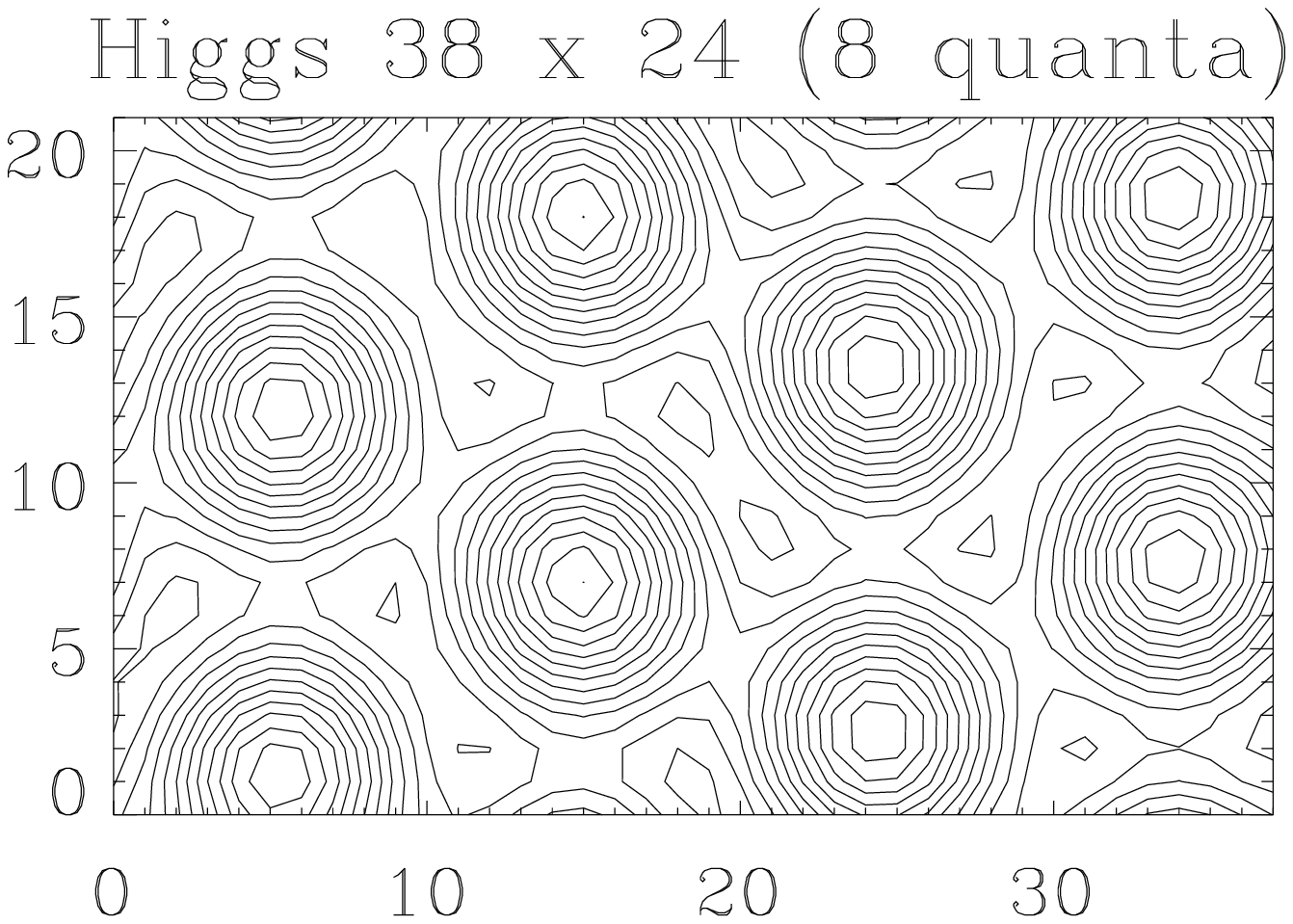}%
}

\vspace*{0.7cm} 

\caption[a]{Left: The tree-level phase diagram of the electroweak
theory in a magnetic field. In the superconductor analogy, 
symmetric phase $\leftrightarrow$ normal phase, 
AO phase $\leftrightarrow$ Abrikosov lattice, 
Higgs phase $\leftrightarrow$ Meissner phase.
Right: The Higgs profile in the 
AO phase\cite{bext2}.
At the point of the minima, the Higgs vev is only $\sim 10...20$\%
smaller than elsewhere.\la{tree_h}}
\end{figure}
%%%%%%%%%%%%%%%%%%%%%%%%%%%%%%%%%%%%

The tree-level phase diagram of the theory in \eq\nr{ewaction}
is shown in \fig\ref{tree_h}. 
This is quite similar to that of the GL model. 
For $m_H>m_Z$ the ground state
solution of the equations of motion is inhomogeneous in 
a certain range of $H_Y$. This Ambj{\o}rn-Olesen (AO) phase\cite{ao} 
is the analogue of the Abrikosov vortex 
lattice of superconductors. There are some differences, 
as well: for instance, the Higgs phase is not really
a Meissner phase, as at low temperatures and fields, 
the magnetic field can pass through the system in a homogeneous
configuration. Another notable difference is that the 
``vortices'' (see \fig\ref{tree_h}(right)) are not
topological objects 
in the same sense as in superconductors, as the Higgs vev
does not vanish at the core of the profile. 

For future reference, let us recall one way of understanding
the appearance of the ``instability'' leading to the AO-phase. 
The point is that at tree-level, there are charged excitations
in both phases of the system
which can be arbitrarily light close enough to the phase 
transition. In the presence of a magnetic field, the corresponding
energies behave as Landau levels. It can then happen 
that some excitations become essentially ``tachyonic'', 
leading to an instability: 
e.g. in the broken phase for large $H_\rmi{EM}$, 
\be
m_{W,\rmi{eff}}^2 = m_W^2 - e H_\rmi{EM} < 0.  \la{constraint}
\ee

\section{The Electroweak Theory in $H_\rmi{ext}\neq 0$ with Fluctuations}

In order to include systematically
the effects of fluctuations, we have studied
the system in \eq\nr{ewaction} with lattice simulations\cite{bext}.
We refer there for the details of the simulations, 
as well as for the justification of the following main result: 
for the values of $H_\rmi{ext}$ studied, we have {\em not observed} 
the AO phase, nor any phase transition at all for $m_H > m_Z$!  
Let us discuss here to what extent we can now understand such a 
contrast with the high-$T_c$ behaviour.

%%%%%%%%%%%%%%%%%%%%%%%%%%%%%%%%% FIGURE
\begin{figure}
 
\vspace*{-0.5cm}

\epsfxsize=6cm
\centerline{\epsffile{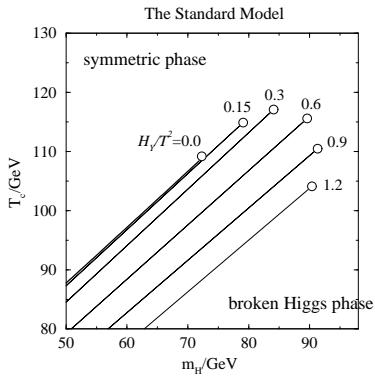}}
 
\vspace*{-3.0cm}

\caption[a]{The non-perturbative phase diagram of the electroweak
theory in a magnetic field (no errorbars shown). 
A solid line indicates a 1st order transition, 
and an open circle a 2nd order endpoint. Based on ref.\cite{bext} and
the preliminary results of ref.\cite{bext2}.\la{fig:res}}
\end{figure}
%%%%%%%%%%%%%%%%%%%%%%%%%%%%%%%%%%%%

For small values of $H_Y$, the discrepancy can be understood as being 
due to SU(2) confinement. For instance, the $W$ is always massive
in contrast to perturbation theory, so that \eq\nr{constraint} 
cannot be satisfied for arbitrarily small $H_\rmi{EM}$. It is 
however difficult to turn this argument into a quantitative one. 

Another way to express the issue is that the only gauge-invariant
degrees of freedom which can become massless are a neutral scalar 
(the Higgs), and the photon\cite{su2u1}. Close to the endpoint
(see \fig\ref{fig:res}), the system can thus be non-perturbatively
described by an effective theory of the form\cite{endpoint} ($\phi\in\RR$) 
\be
{\cal L} = \fr14 F_{ij}F_{ij} + 
\fr12 (\partial_i\phi)^2 +  h \phi + 
\fr12 m^2 \phi^2 + \fr14 \lambda \phi^4 + 
\gamma_1 \phi F_{ij}F_{ij} + ...\, .  \la{fineff}
\ee
However, in this theory there are no charged excitations, hence no 
Landau levels and instabilities, unlike at tree-level!
 
On the other hand, the effective theory in \eq\nr{fineff}
can in principle break down for very large fields, and also far 
away from the endpoint, and one may ask what happens then?
It is here that the case of superconductors again becomes relevant. 
As discussed at the end of Sec.~3, it might be that even in superconductors
some extra structure such as layers is needed in order to have
a vortex phase and the associated 1st order transition. If so, 
then it is unlikely that there would be any remnant of the
AO phase in the fluctuating electroweak system even at large $H_Y$.
If no layers are needed, on the contrary, there just might be one.

\section{Conclusions}

It appears that even if there is an external magnetic field
present, the SM electroweak transition terminates at 
$m_H\lsim 90$ GeV, and above that there is no structure at all, 
see \fig\ref{fig:res}. In particular, the Ambj{\o}rn-Olesen phase 
seems not to be realized at realistic magnetic fields. Thus an 
electroweak phase transition within the SM does not leave a cosmological 
remnant. An interesting theoretical open issue is still what
happens at very large magnetic field strengths --- a 
question which involves quite intriguing analogies also with 
the behaviour of experimentally accessible high-$T_c$ superconductors. 

\section*{Acknowledgments}

I thank  K. Kajantie, 
T. Neuhaus, P. Pennanen, A. Rajantie, K. Rummukainen, M. Shaposhnikov 
and M. Tsypin for collaboration and discussions
on various topics mentioned in this talk.  
This work was partly supported by the 
TMR network {\em Finite Temperature Phase
Transitions in Particle Physics}, EU contract no.\ FMRX-CT97-0122.

%%%%%%%%%%%%%%%%%%%%%%%%%%%%%%%%%%%%%%%%%%%%%%%%%%%%%%%%%%%%%%%%%%%%%%%%
%%%%%%%%%%%%%%%%%%%%%%%%%%%%%%%%%%%%%%%%%%%%%%%%%%%%%%%%%%%%%%%%%%%%%%%%

\end{document}